\documentclass[sigconf]{acmart}  

\usepackage{booktabs} 

\setcopyright{rightsretained}

\acmConference[WORKS'17]{WORKS 2017 Workshop}{November 2017}{Denver, CO,
USA}
\acmYear{2017}
\copyrightyear{2017}

\begin{document}
\title{revisit: a Workflow Tool for Data Science}

\author{Norman Matloff}
\affiliation{%
  \institution{University of California, Davis}
  \city{Davis} 
  \state{CA} 
  \postcode{95616}
}
\email{matloff@cs.ucdavis.edu}

\author{Reed Davis}
\affiliation{%
  \city{Sunnyvale} 
  \state{CA} 
}
\email{rdavis2468@gmail.com}

\author{Laurel Beckett}
\affiliation{%
  \institution{University of California, Davis}
  \city{Davis} 
  \state{CA} 
  \postcode{95616}
}
\email{labeckett@ucdavis.edu}

\author{Paul Thompson}
\affiliation{%
  \institution{Sanford Research}
  \city{Sioux Falls} 
  \state{SD} 
  \postcode{57104}
}
\email{Paul.Thompson@sanfordhealth.org}

\renewcommand{\shortauthors}{N. Matloff et al.}

\begin{abstract}

In recent years there has been widespread concern in the scientific
community over a {\it reproducibility crisis}.  Among the major causes
that have been identified is statistical: In many scientific research
the statistical analysis (including data preparation) suffers from a
lack of transparency and methodological problems, major obstructions to 
reproducibility.  The {\bf revisit} package aims toward remedying this problem, 
by generating a ``software paper trail'' of the statistical operations 
applied to a dataset.  This record can be ``replayed'' for verification
purposes, as well as be modified to enable alternative analyses.  The
software also issues warnings of certain kinds of potential errors in
statistical methodology, again related to the reproducibility issue.

\end{abstract}



\maketitle

\section{The Reproducibility Crisis}

In recent years, scientists, especially those who run academic journals
or fund research projects, have been greatly concerned about lack of
{\it reproducibility} of research.  A study performed by one research group,
with certain findings, is then attempted by another group, with
different findings.  In addition, there is a related problem, lack of
{\it transparency}. In reading a paper reporting on certain research, it
is often not clear exactly what procedures the authors used.  

The problem is considered by many to have reached crisis stage
\cite{Ioannidis-2005-696-701} \cite{baker}.

\subsection{The Statisical Aspects}

Though many problems of reproducibility are due to experimental issues
such as subtle differences in procedures from one laboratory to another,
much of the concern is statistical in nature. As noted in \cite{baker}
(emphasis added):

\begin{quote}

The survey asked scientists what led to problems in reproducibility.
More than 60\% of respondents said that each of two factors --- pressure
to publish and selective reporting --- always or often contributed. More
than half pointed to insufficient replication in the lab, poor oversight
or low statistical power.

Respondents were asked to rate 11 different approaches to improving
reproducibility in science, and all got ringing endorsements. Nearly
90\% --- more than 1,000 people --- ticked ``More robust experimental
design,'' ``better statistics''...

\end{quote}

It should be noted that in 2016 the American Statistical Association, in
its first foundational public policy statement in its 177-year history,
issued stern guidance on the overuse and misinterpretation of p-values
\cite{ronw}.  Though stopping short of recommending an outright ban, one
psychology journal did exactly that \cite{siegfried}.  The ASA statement
noted that its recommendation was being made, {\it inter alia}, to ``...
inform the growing emphasis on reproducibility of science research,''
and recommended that researchers move more to using confidence
intervals and other approaches, instead of significance tests.

Just as important is the transparency issue.  What steps did the
researchers take during data preparation?  Did they check for outliers?
If so, what methods did they use for this, and on what basis did they
decide to include or exclude an outlying data point?  In using
statistical software libraries, what options did they use?  For that
matter, in view of the fact that such libraries can change over time in
subtle ways, what versions of the libraries did they use?

Obtaining the same result from a given set of data sounds obvious and
trivial, but there are a number of reasons why this may fail to happen.
First, certain types of analyses are not closed-form but rather are
iterative and approximating.  Unless the same convergence criteria,
start values, and step size levels are used, it is entirely possible to
get different outcomes. This is particularly true in cases in which
the results surface is relatively flat or the objective function is
nonconvex.

 Second, analyses may be done in an interactive manner, and thus the
tracking of the exact processes involved can sometimes be difficult.
When interactive methods are used, it is possible that steps are
forgotten, or that steps are done in different orders.

The well-known case of Potti {\it et al} is an instructive example of
why there is so much concern.  A link between patients' gene expression
and their response to cancer treatment had been reported in multiple
papers in highly regarded journals, leading to clinical trials based on
this purported link.  The original data were analyzed by a physician who
was not well trained in proper data analysis, proper data storage, or
proper use of training and validation samples \cite{Baggerly-2011-1}
\cite{Reich-Nature-2011}.  The analysis methods were not publicly
available, but intensive detective work by two biostatisticians at MD
Anderson with the original (publicly available) dataset, over a
year-long period, cast doubt on the findings, and eventually led to the
conclusion that the results were incorrect, due to, to start with,
extreme carelessness, and much more.  A number of errors were found,
including changes in the version of the main data analysis tool
\cite{Baggerly-2011-1} \cite{hayden}.  Other errors included questions
about the integrity of the sample \cite{Baggerly-2011-1}. In many
studies, the sample being examined changes, and unless care is taken to
``freeze'' the sample, different analyses may be made from different
samples.

The trials were discontinued and lawsuits followed, amid much publicity,
including coverage on the CBS news show {\it 60 Minutes}.  While the
extent of publicity was highly unusual, the lack of transparency about
data handling, analysis methods, and interpretation of findings was
commonplace, as seen by publication of results in major journals.  As a
result, proposals to the National Institutes of Health are now required
to include a section on rigor and reproducibility, and grant reviews
must assess this component \cite{nih}.

Reproducing analyses have revealed problems and potential alternative
interpretations in fields other than medicine. Two noted economists
published a data analysis suggesting that if national debt exceeds 90\%
of Gross Domestic Product (GDP), the rate of economic growth will slow
dramatically \cite{pollin}. International monetary groups cited these
results in policy decisions. A graduate student, taking the publicly
available data used in the paper, tried to replicate their results and
could not; the economists' analyses, carried out in Excel, were
apparently affected both by data errors and by specific choices of
analysis method that could be influenced by outliers.  Reproducing their
analysis and considering alternative approaches, on a corrected dataset,
allowed for a much more nuanced discussion of the relationship between
debt and growth, with multiple possible interpretations of the data. 

Many of the problems with complete and fully accurate reproducibility
of results occur because the analysis is incompletely documented or
tracked. Analysis of data begins with data management. Data are
obtained from sources (data collection tools). The data are filtered
(i.e., invalid values are removed or corrected, incorrect cases are
sometimes removed if they were incorrectly added). Data management is
a key step in the process and must be carefully documented for later
checking and examination.

Such transparency does not preclude a rebuttal by the original authors
of a paper whose results have been questioned \cite{pollin}.  On the
contrary, transparency is what enables such interaction, with science
being the winner.

\section{The revisit Software Tool: Overview}

Statistical software tools must thus be developed to address such
problems \cite{Peng-2011-1226-1227}.  Moreover, reproducibility
essentially implies a scripting approach, rather than a GUI; it is
difficult if not impossible to record mouse clicks in a manner that is
``replayable'' across platforms.  Scripting code can be inspected,
transfered to others, used on more than one project, and modified
easily. It also functions as the memory of the project
\cite{Thompson-2016-2}.

The {\bf revisit} package, available at {\it
https://github.com/matloff/revisit}, addresses the reproducibility issue
from a workflow perspective, both in terms of transparency and in
statistical quality of the work.

In one sense, the package might be said to enable a {\it statistical
audit}, allowing users to check the statistical analyses of the original
authors of a study, but it really is much more.  In our referring to
``users'' below, keep in mind that this could mean various kinds of
people, such as:

\begin{itemize}

\item {\it The various authors of the study}, during the period when the
study is being conducted.  The package will facilitate collaboration
among the authors during that time.

\item {\it Reviewers of a manuscript on the study}, presented for possible
publication.  The package will facilitate the reviewers' checking of the
statistical analyses in the paper, not only verifying the steps but,
even more importantly, allowing the reviewers to explore alternative
analyses.

\item {\it Other scientists}, who are reading a published paper on the
study.  The package will facilitate these scientists to also explore
various alternative analyses.

\end{itemize}

The package has two main aspects:

\begin{itemize}

\item [(a)] It makes it easier and more convenient for the user to
explore the effects of various changes that might be made to the
analyses.  The package facilitates

   \begin{itemize}

   \item  replaying the analysis;

   \item  changing it; and

   \item  recording changed versions.

   \end{itemize}

\item [(b)] The package attempts to spot possibly troublesome statistical
situations, and issues advice and warnings, in light of concerns that
too many ``false positives'' are being reported in published research.
For example, the package may:\footnote{Some of these features are not
yet implemented. Features are being added on an ongoing basis.}

   \begin{itemize}

   \item Point out that although a certain p-value is small, it may not
   correspond to an effect of practical importance.

   \item Point out that a ``nonsignificant'' value corresponds to a
   confidence interval containing both large positive and large negative
   values, so that the sample size $n$ is too small for a ``no significant 
   difference'' finding.

   \item Suggest that the user employ a multiple inference procedure,
   say Bonferroni's approach or a more advanced method
   \cite{matloffreg}, to help avoid finding spurious
   correlations.\footnote{See \cite{aschwanden} for a humorous but
   highly illuminating example.}

   \item Detect the presence of highly influential outliers, and suggest
   that a robust method, e.g.\ quantile regression, be used.

   \item Detect evidence of possible overfitting.

   \item Etc.

   \end{itemize}

\end{itemize}

More specifically, the user might go through the following thought
processes, and take action using the facilities in the package:

\begin{itemize}

\item  The user thinks of questions involving alternate scenarios.
What, for instance, would occur if one were to more aggressively weed
out outliers, or use outlier-resistant methods?  How would the results
change?  What if different predictor variables were used, or squared and
interaction terms added?  How valid are the models and assumptions?
What about entirely different statistical approaches?

\item  In exploring such questions, the user will modify the original
code, producing at least one new version of the code, typical several.
Say for instance the user is considering making two changes to the
original analysis, one to possibly use outlier-resistant methods and
another to use multiple-inference procedures. That potentially sets up
four different versions.  The {\bf revisit} package facilitates this,
making it easier for the user to make changes, try them out and record
them into different {\it branches} of the code, similar to GitHub.  In
other words, the package facilitates exploration of alternative
analyses.

\item  In addition, the user may wish to share the results of her
exploration of alternate analyses of the data with others.  Since each
of her branches is conveniently packaged into a separate file, she then
simply sends the files to the other researchers.  The package allows the
latter to easily ``replay'' the analyses, and they in turn may use the
package to produce further branches.

\end{itemize}

\section{Previous Work} 

Due to the heavy interest in the reproducibility issue in recent years,
a number of efforts have been made in the software realm. 

One direction such efforts have taken is the development of software to
facilitate integration of statistical analysis performed through the R
programming language with research reports and papers. One of the CRAN
Task Views on the R Project site is devoted to this issue
\cite{crantask}, as is a book \cite{gandrud}.

Other projects have been aimed at increasing transparency.  For
instance, \cite{karthik} notes,

\begin{quote}

Reproducibility is the hallmark of good science. Maintaining a high
degree of transparency in scientific reporting is essential not just for
gaining trust and credibility within the scientific community but also
for facilitating the development of new ideas. Sharing data and computer
code associated with publications is becoming increasingly common,
motivated partly in response to data deposition requirements from
journals and mandates from funders...

\end{quote}

More directly addressing the workflow issue is \cite{bertram}, in a
general scientific context.  A similar goal, aimed at parallel
computation but still in a general scientific context is \cite{faceit}.
Another example, in the context of computational harmonic analysis, is
presented in \cite{Donoho-2009-8-18}.

Our {\bf revisit} package takes a different path.  Though it too is
workflow-oriented, it is specific to statistical applications.

\section{The revisit Software Tool: Details}

The software is written in R. The author of the original scientific
research is assumed to do all data/statistical analysis in R.

Both text-based and graphical (GUI) interfaces of {\bf revisit} are
available. The GUI uses RStudio {\it add-ins} \cite{rstudio}. The
text-based version provides more flexiblity, while the GUI provides
convenience.

Our first example here uses the famous Pima diabetes study at the UCI
data repository \cite{pima}. The following table shows the 9 variables
in the data file {\bf pima.txt}, followed by their descriptions:

\bigskip

\begin{tabular}{|l|l|}
\hline
Variable     &   Description \\
\hline
NPreg   & Number of times pregnant \\
\hline
Gluc    & Glucose  \\
\hline
BP      & Diastolic blood pressure  \\
\hline
Thick   & Triceps skin fold thickness  \\
\hline
Insul   & Serum insulin  \\
\hline
BMI     & Body mass index  \\
\hline
Genet   & Diabetes pedigree function \\
\hline
Age     & Age  \\
\hline
Diab    & Diabetic class variable (0 or 1) \\
\hline
\end{tabular}

\bigskip

Say Scientist X is a researcher authoring the study.  The idea of {\bf
revisit} is that, during the course of the study, he would record all
data operations in an R file, say {\bf pima.R}.  (The data must be
included too.)

\begin{figure*}
\includegraphics[width=6in]{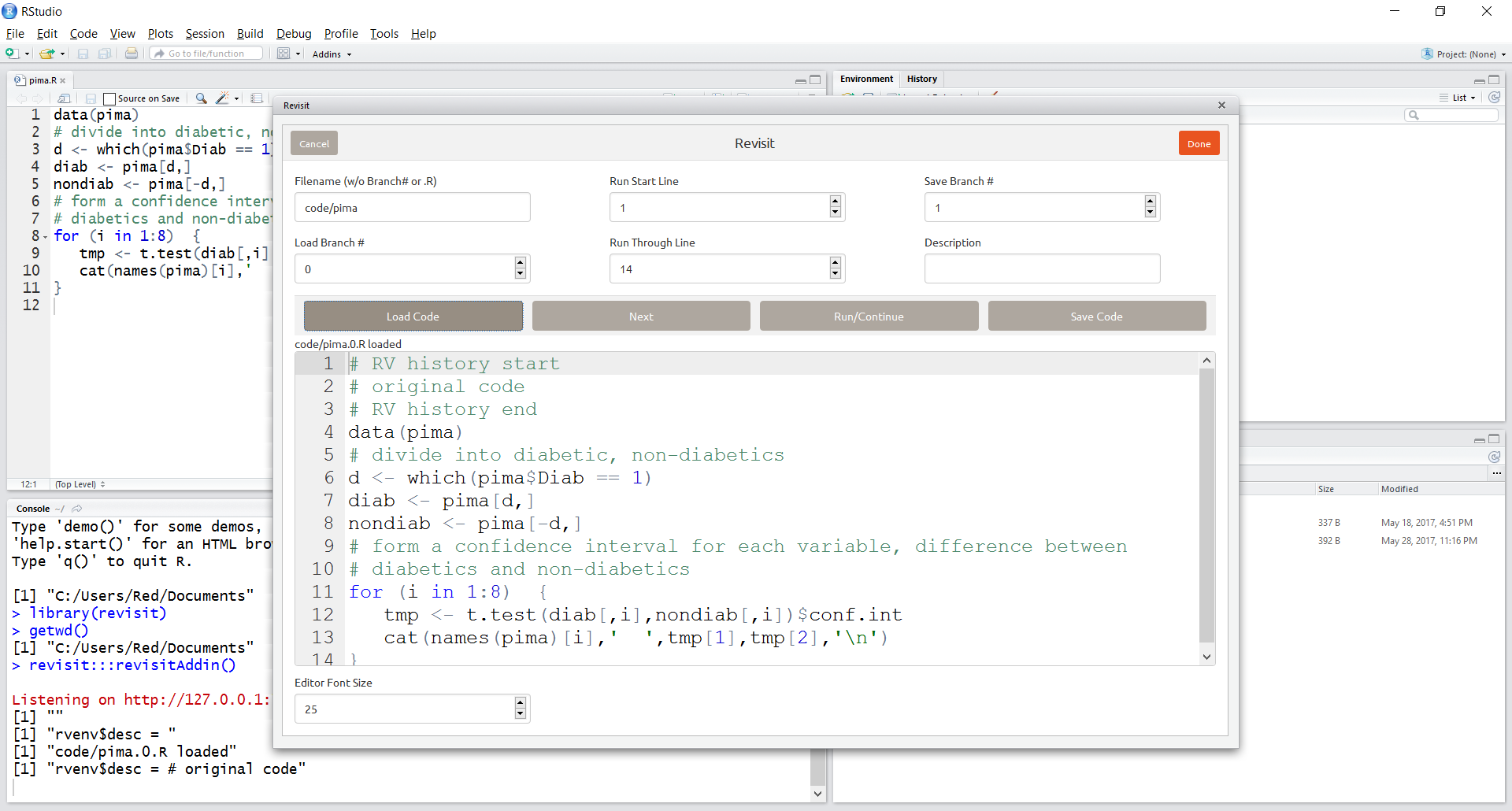}
\caption{Opening screen}
\label{openingscreen}
\end{figure*}

\begin{figure*}
\includegraphics[width=6in]{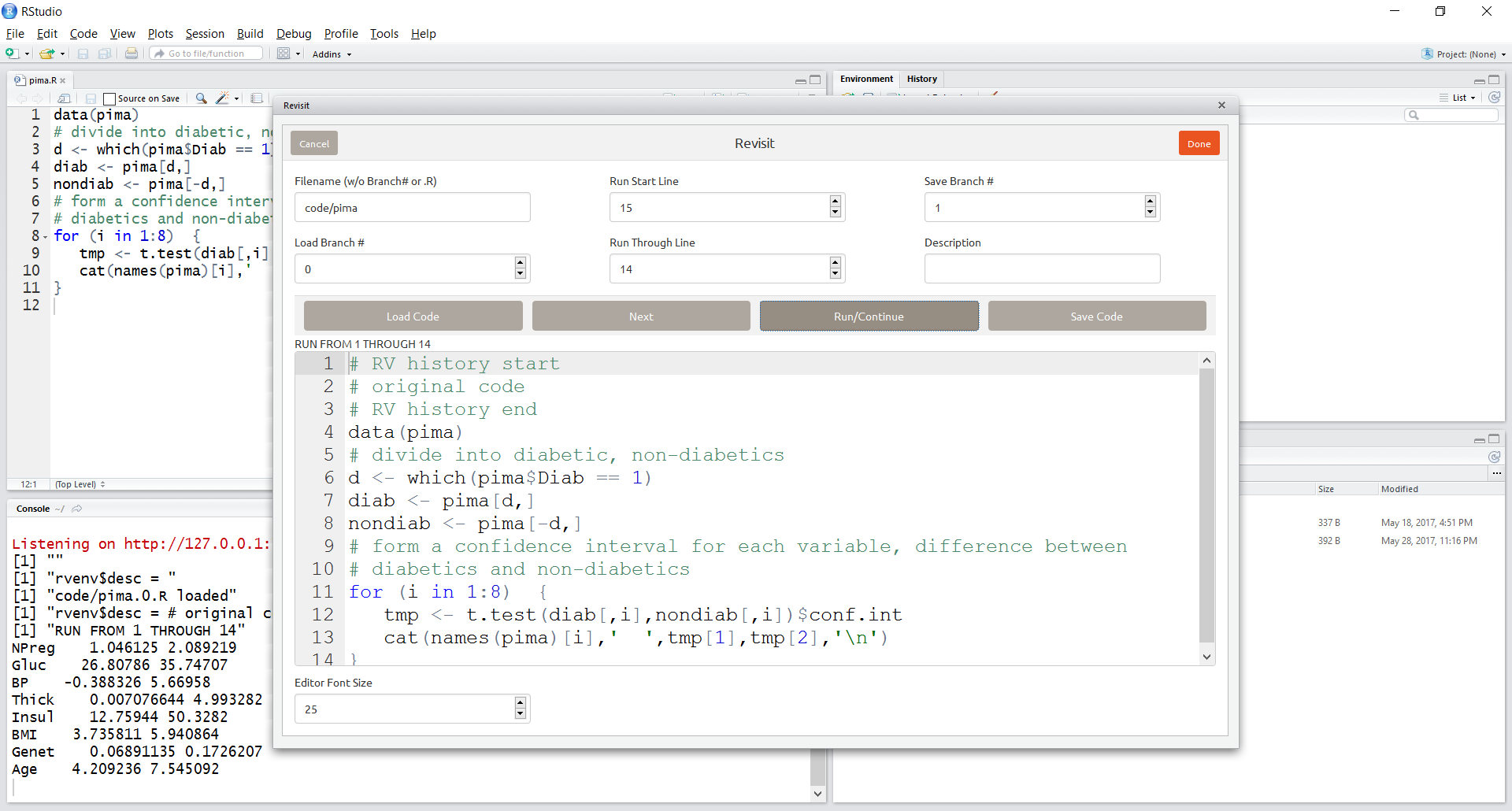}
\caption{After first run}
\label{after1strun}
\end{figure*}

Now suppose the study has already been published , and Scientist Y is
interested in it.  (Other possible roles for Y might be, as noted
earlier, as a member of the research team during the course of the
study, or as a reviewer of a manuscript submitted for publication.) The
point is that {\bf revisit} will (a) enable Y to confirm X's statistical
results, and (b) explore alternative statistical approaches.  Here is
how:

First, Y would load {\bf pima.R}.  The {\bf revisit} screen would then
look like Figure \ref{openingscreen}.  Y will then see X's code, and
will now be free to run and/or modify it.  To replay X's code without
modfication, Y clicks Run/Continue.\footnote{The package and GUI
interface have various convenience features.  For instance, the user can
set the starting line, and also opt to run through only a specified
line, rather than running the entire code.  See the package
documentation for details.} The new screen is shown in
Figure \ref{after1strun}, including the output from the run.

\begin{figure*}
\includegraphics[width=6in]{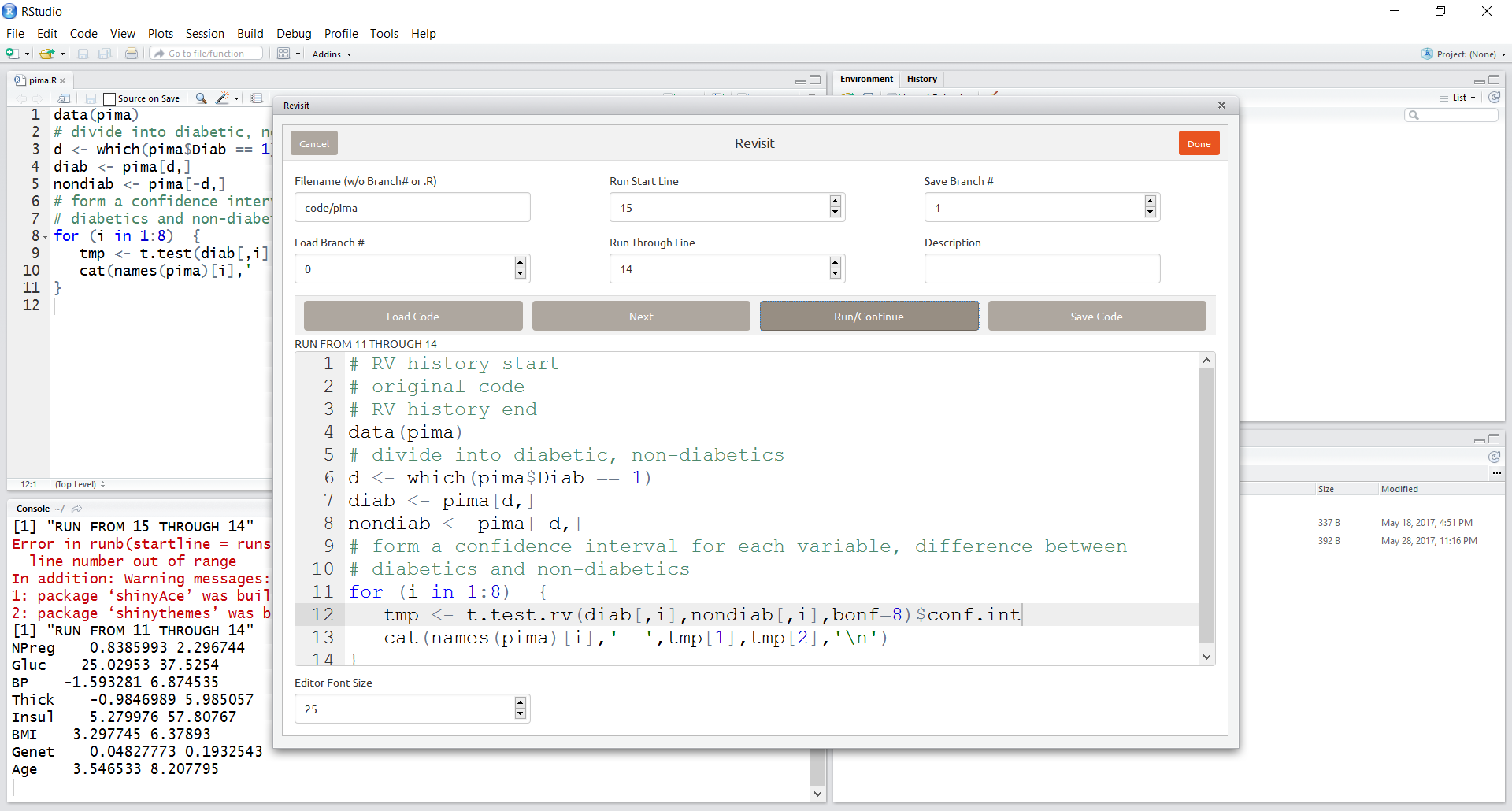}
\caption{After Bonferroni}
\label{postbonf}
\end{figure*}

At this point, Y might feel that forming eight confidence intervals
risks ``accidental'' findings, effects arising from pure chance rather
than substantial differences between the diabetic and non-diabetic
groups.  So, Y might ask, ``How would the results change if we were to
use Bonferroni's method here?''  So Y changes line 12 to use {\bf
revisit}'s own function, which employs the Bonferroni method, with
number of comparisons equal to 8.  The result is depicted in Figure
\ref{postbonf}.  Ah, the confidence intervals did indeed get wider, as
expected with the $\alpha$ adjustment, in line with statistical
fairness.  Views by domain experts may change as a result.  

Now, Y may wish to do further exploration of alternative analyses, and
may also wish to share the various versions of the code with others. So,
a key feature of {\bf revisit} is the ability to save various versions
of the code, which Y will probably wish to do as she explores more and
more modifications to X's code.

As noted before, in addition to formal statistical analysis, another
aspect of reproducibility is transparency of the data preparation
process (often called {\it data cleaning}, {\it data wrangling} or {\it
data munging}). For instance, the presence of outliers can have a
serious impact on the outcome of one's statistical analysis, so Y might
ask, ``What did X do about possible outliers?''  In this case, she would
discover from X's code that X had done nothing at all in this regard,
and Y may wish to explore this avenue.

As a crude measure, Y could find the range of each of the variables, say
by running the code

\bigskip
\begin{verbatim}
print(apply(pima[,1:8],2,range))
\end{verbatim}

\begin{figure*}
\includegraphics[width=6in]{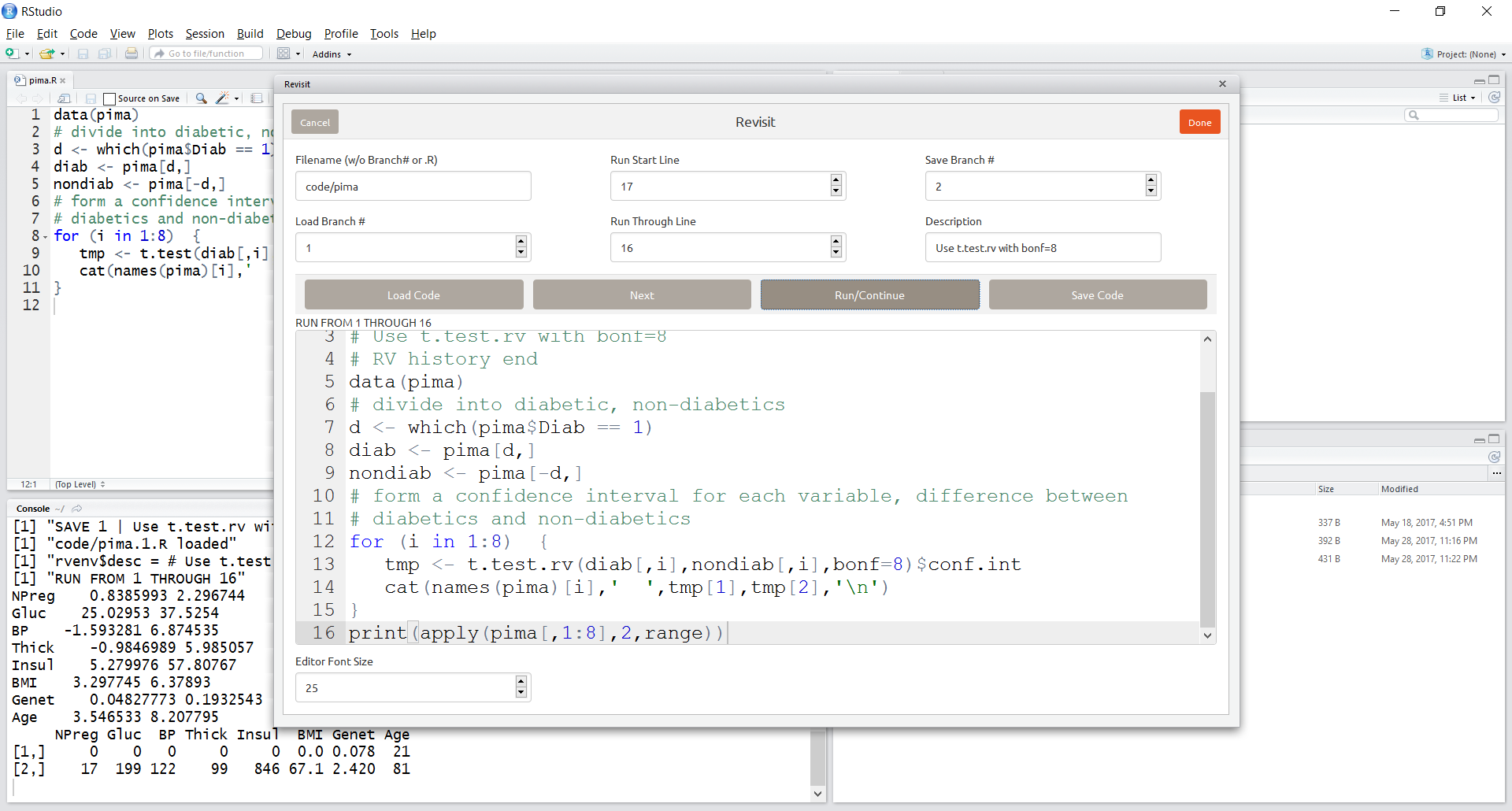}
\caption{Outlier hunt}
\label{outliers}
\end{figure*}

\bigskip
\noindent
The actual operations for this would differ between the GUI and
text-based versions of {\bf revisit}.  In the GUI version, Y would
temporarily add the above line to the end of the current code, then run,
resulting in Figure \ref{outliers}.

The top and bottom rows of the output are the minimum and maximum values
of the given variable.  Those 0s are troubling. How can variables such
as Glucose and BMI be 0?  The descriptions of the variables above
suggest that the 0s for the variables Gluc, BP, Thick, Insul, and BMI
actually represent missing values. So Y can set the 0s to missing with
the following statements:

\bigskip
\begin{verbatim}
pima$Gluc[pima$Gluc == 0] <- NA
pima$BP[pima$BP == 0] <- NA
pima$Thick[pima$Thick == 0] <- NA
pima$Insul[pima$Insul == 0] <- NA
pima$BMI[pima$BMI == 0] <- NA
\end{verbatim}

\bigskip
\noindent
This too becomes part of the code record, which Y will likely want to
save, in a new version.  In fact, Y may wish to develop two parallel
branches from this point, one with and one with outlier removal.

\section{revisit: Further ``Statistical Audit'' Features}

As with income tax preparation software that gives advice, say warning
that a certain deduction may be questionable, {\bf revisit} comments on
possible misuses of statistical methodology.

Consider the MovieLens data \cite{ml}, a very popular example
dataset in recommender systems research.  Suppose we are interested in
the question of whether user demographics has a relation to user
ratings.  As a first-level analysis, we might try a linear regression
model, predicting rating from age and gender.  A standard R analysis
would go something like, 

\bigskip
\begin{verbatim}
lmout <- lm(usermeans ~ uu[,2] + uu[,3]) 
print(summary(lmout))  # get estimates, p-values etc. 
\end{verbatim}

\bigskip
\noindent
with (partial) output

\bigskip
\begin{verbatim}
Coefficients:
             Estimate Std. Error t value Pr(>|t|)    
(Intercept) 3.4725821  0.0482655  71.947  < 2e-16 ***
uu[, 2]     0.0033891  0.0011860   2.858  0.00436 ** 
uu[, 3]     0.0002862  0.0318670   0.009  0.99284    
\end{verbatim}

\bigskip
\noindent
This reports that age (though not gender) has a ``highly significant''
positive relationship with rating.  But instead of calling R's {\bf
summary()} function, Y could call the corresponding facility in {\bf
revisit}:

\bigskip
\begin{verbatim}
> coef.rv(lmout)
          est.         left       right      p-val warning
1 3.4725821093  3.357035467 3.588128752 0.00000000        
2 0.0033891042  0.000549889 0.006228319 0.01280424     X
3 0.0002862087 -0.076002838 0.076575255 1.00000000        
\end{verbatim}

\bigskip
\noindent
Note the X in the ``warning'' column.  The estimated age coefficient
here, about 0.0034, is tiny; a 10-year difference in age corresponds to a
difference in mean rating of only about 0.034, minuscule for ratings in
the range of 1 to 5. This ``highly significant'' result is likely of no
practical interest.

As noted earlier, the misuse of p-values has been cited as a factor
leading to the reproducibility crisis. Our software attempts to flag
problems in this regard.

\section{Future Work}

A number of further ``statistical audit'' features are planned, such as:

\begin{itemize}

\item Further options for multiple-inference procedures \cite{hsu}. In
keeping with the theme of reducing use of p-values, emphasis will be
placed on procedures that involve confidence intervals.
Some {\it postinference} methods \cite{berk2013} will also be
considered.

\item Further options for outlier detection and for outlier-robust 
methods \cite{matloffreg}.

\item Further assistance for moving toward inference based on confidence
intervals rather than significance tests.  

For example, the analysis of {\it contingency tables} is usually based
solely on significance tests.  The standard R function for this, {\bf
loglin()}, will go beyond this only if one invokes an option to obtain
point estimates.  But even then, standard errors are not provided, so 
that confidence intervals cannot be formed. A workaround is possible, by
exploiting the fact that Poisson inputs to the table are conditionally
multinomial \cite{christensen}; this enables use of the R generalized
linear model function {\bf glm()}, which does provide standard errors.
A wrapper for this procedure will be developed.

\end{itemize}

\section{Conclusions}

The {\bf revisit} package addresses the statistics/data management
aspects of the reproducibility problem by enabling users to replay the
data analysis of a research project, and conveniently explore
alternative analyses. These modified analyses can easily be shared with
others, facilitating not only post-research discussion but also
collaboration among researchers during the course of a project.

Furthermore, the package acts as a ``statistical audit,'' warning of
potential trouble spots and suggesting improvements.  This goes to the
heart of many statistical problems that have been identified as
contributing to the reproducibility crisis, particularly the overuse of
significance tests and the lack of use of multiple-inference methods.

\bibliographystyle{ACM-Reference-Format}
\bibliography{works17.bib} 


\end{document}